# Comment on "Isotope Dependence of the Spin Gap in YBa$_2$Cu$_4$O$_8$ as Determined by Cu NQR Relaxation"

Underdoped HTS cuprates are characterised by a normal-state pseudogap in the quasiparticle spectrum extending well above T$_c$ as first evidenced by a depression in Knight shift and electronic entropy and more recently by direct observation using ARPES or tunnelling spectroscopy [1]. The Knight shift, K$_s$, is proportional to the static, long-wavelength (**q**=0) spin susceptibility. There is also a widespread belief that the cuprates display a separate gap in the spin excitation spectrum which opens up at T*, the position of the peak in 1/T$_1$T [2]. Here T$_1$ is the copper spin-lattice relaxation time which is strongly weighted towards **q**=($\pi,\pi$). In a recent Letter, Raffa et al. [3] showed from high-resolution NQR studies on YBa$_2$Cu$_4$O$_8$ the existence of an oxygen isotope effect in the temperature dependence of 1/T$_1$T. The similar values of the isotope effect exponents $\alpha_{Tc}$ = 0.056 and $\alpha_{T*}$ = 0.061 led the authors to infer a common origin for the superconducting and spin gaps. Earlier [4], we had found no change with oxygen isotope in the temperature dependence of high-precision $^{89}$Y Knight shift magic-angle-spinning (MAS) data and concluded the absence of an isotope effect in the pseudogap energy E$_g$ within the margin $\alpha_{Eg}$ < 0.01. Raffa et al state that this confirms a distinction has to be made between the behaviour of the spin susceptibility at **q**=0 and **q**=($\pi,\pi$).

In fact, we have argued that as 1/$^{63}$T$_1$ (i.e. excluding the 1/T enhancement) is found to have *identical* temperature dependence to K$_s$ they both have the same underlying energy scale E$_g$ and the spin gap and the pseudogap are one and the same [5]. This concurs with the view of Loram et al. [6] from heat capacity that the normal-state gap is the same quasiparticle gap observed for all **q**. Our $^{89}$K$_s$ isotope effect data showed a small but possibly significant isotope shift below 170K. We can now confirm this as a systematic effect. In Fig. 1 we plot 1/T$_1$ from the data of Raffa et al. together with our MAS $^{89}$K$_s$ data. The remarkable correspondence between the two confirms that they possess the same underlying energy scale and represent the same gap in the quasiparticle spectrum. The inset shows the isotope-induced difference in $^{89}$K$_s$ (circles) and in 1/$^{63}$T$_1$ (dots). The two are rather consistent, each showing a positive excursion below 170K and roughly compatible with the isotope-shifted curve (solid line) assuming that $\alpha_{Eg} = \alpha_{T*} = \alpha_{Tc}$ = 0.056. However, even if $\alpha_{Eg} \approx \alpha_{Tc}$ this may well be mere coincidence: the precise agreement between the relaxation rate and the shift shows that the pseudogap is unaffected by magnetic field (zero-field NQR compared with $^{89}$Y NMR at 11.74T) thus suggesting that the pseudogap and the pairing correlations are of different origin. Moreover, there is no significant isotope shift in the Knight shift above 170K. Finally, we show (triangles) the change in $^{89}$K$_s$ for $^{65}$Cu relative to $^{63}$Cu in YBa$_2$Cu$_4$O$_8$ [7]. This also shows a small isotope effect below 170K but the excursion is negative while the exponent for the copper isotope shift in T$_c$ is positive ($\alpha_{Tc}$ = +0.09 ±0.01, see also ref [8]). We reaffirm that the spin gap and pseudogap are the same quasiparticle gap with an energy scale that climbs to J at zero doping and having a distinct origin from the superconducting gap.

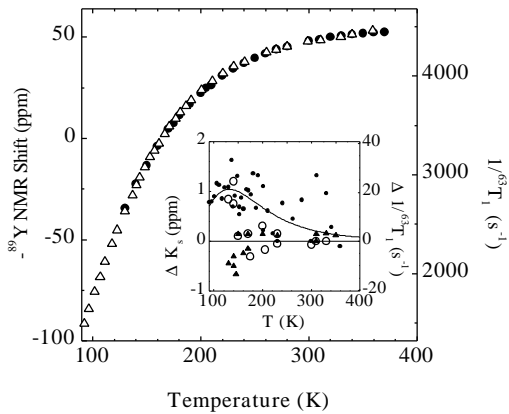

Fig.1. The T-dependence of $^{89}$Y Knight shift (●) and $^{63}$R$_1$ = 1/$^{63}$T$_1$ (△) for Y-124. Inset: the O-isotope shift in K$_s$ (○) and in $^{63}$R$_1$ scaled by 1/17 (●) and the Cu-isotope shift in K$_s$ (△).

J.L. Tallon, G.V.M. Williams and D. J. Pringle*
  N.Z. Institute for Industrial Research
  P.O. Box 31310, Lower Hutt, New Zealand
  *School of Chemical and Physical Sciences
  Victoria University,
  P.O. Box 600, Wellington, New Zealand